\documentclass[10pt,sigconf,letterpaper]{acmart}

\usepackage{array}
\usepackage[english]{babel}
\usepackage[utf8]{inputenc}
\usepackage{blindtext}
\usepackage{multirow}
\usepackage[absolute,showboxes]{textpos}

\usepackage{booktabs}

\usepackage{cleveref}
\crefformat{section}{\S#2#1#3}
\crefformat{subsection}{\S#2#1#3}
\crefformat{subsubsection}{\S#2#1#3}
\crefrangeformat{section}{\S\S#3#1#4 to~#5#2#6}
\crefmultiformat{section}{\S#2#1#3}{ and~\S#2#1#3}{, \S#2#1#3}{ and~\S#2#1#3}

\usepackage{xspace}
\usepackage{graphicx}
\usepackage{balance}
\usepackage{layouts}

\let\myRef\ref
\renewcommand\ref{\unskip~\myRef}
\let\myCite\cite
\renewcommand\cite{\unskip~\myCite}

\newcommand{\ie}{i.e., \@}
\newcommand{\eg}{e.g., \@}

\newcommand{\perc}{\,\%\xspace}

\newcommand{\etal}{et al.\xspace}
\newcommand{\one}{(1)~}
\newcommand{\two}{(2)~}
\newcommand{\three}{(3)~}
\newcommand{\four}{(4)~}
\newcommand{\five}{(5)~}
\newcommand{\msp}{more-specific prefix\xspace}
\newcommand{\msps}{{\msp}es\xspace}

\newcommand{\mspsS}{MSPs\xspace}
\newcommand{\hsp}{hyper-specific prefix\xspace}
\newcommand{\hsps}{{\hsp}es\xspace}
\newcommand{\hspS}{HSP\xspace}
\newcommand{\hspsS}{HSPs\xspace}

\newcommand{\ra}{RIPE Atlas\xspace}
\setcopyright{none}
\renewcommand\footnotetextcopyrightpermission[1]{}
\acmDOI{}
\acmISBN{}
\acmPrice{}
\renewcommand\footnotetextcopyrightpermission[1]{}
\acmConference{}{}{}

\usepackage{fancyhdr}
\fancypagestyle{plain}{%
   \fancyhf{} %
   \fancyfoot[L]{ACM SIGCOMM Computer Communication Review}%
   \fancyfoot[R]{Volume 52 Issue 2, April 2022}%
}
\fancypagestyle{firstpagestyle}{%
   \fancyhf{} %
   \fancyfoot[L]{ACM SIGCOMM Computer Communication Review}%
   \fancyfoot[R]{Volume 52 Issue 2, April 2022}%
}

\begin{document}

\title{Hyper-Specific Prefixes: \\Gotta Enjoy the Little Things in Interdomain Routing}

\author{Khwaja Zubair Sediqi}
\affiliation{
	\institution{MPI-INF}
}
\email{zsediqi@mpi-inf.mpg.de}

\author{Lars Prehn}
\affiliation{
	\institution{MPI-INF}
}
\email{lprehn@mpi-inf.mpg.de}

\author{Oliver Gasser}
\affiliation{
	\institution{MPI-INF}
}
\email{oliver.gasser@mpi-inf.mpg.de}

\begin{CCSXML}
<ccs2012>
<concept>
<concept_id>10003033.10003039</concept_id>
<concept_desc>Networks~Network protocols</concept_desc>
<concept_significance>500</concept_significance>
</concept>
</ccs2012>
\end{CCSXML}

\ccsdesc[500]{Networks~Network protocols}

\keywords{BGP, routing, hyper-specific prefixes.}

\begin{abstract}

Autonomous Systems (ASes) exchange reachability information between each other using BGP---the de-facto standard inter-AS routing protocol. 
While IPv4 (IPv6) routes more specific than /24 (/48) are commonly filtered (and hence not propagated), route collectors still observe many of them.

In this work, we take a closer look at those ``hyper-specific`` prefixes (HSPs).  In particular, we analyze their prevalence, use cases, and whether operators use them intentionally or accidentally.
While their total number increases over time, most HSPs can only be seen by route collector peers.
Nonetheless, some HSPs can be seen constantly throughout an entire year and propagate widely. 
We find that most HSPs represent (internal) routes to peering infrastructure or are related to address block relocations or blackholing.
While hundreds of operators intentionally add HSPs to well-known routing databases, we observe that many HSPs are possibly accidentally leaked routes.

\end{abstract}

\setlength{\TPHorizModule}{\paperwidth}
\setlength{\TPVertModule}{\paperheight}
\TPMargin{5pt}
\begin{textblock}{0.8}(0.1,0.02)
    \noindent
    \footnotesize
    If you cite this paper, please use the SIGCOMM CCR reference:
    Khwaja Zubair Sediqi, Lars Prehn, Oliver Gasser. 2022.
    Hyper-Specific Prefixes: Gotta Enjoy the Little Things in Interdomain Routing.
    In \textit{ACM SIGCOMM Computer Communication Review, Volume 52, Issue 2, April 2022.}
    ACM, New York, NY, USA, 14 pages. \url{https://doi.org/10.1145/3544912.3544916}
\end{textblock}

\maketitle

\section{Introduction}
\label{sec:introduction}

Autonomous Systems (ASes) use the Border Gateway Protocol (BGP) to announce prefixes to their peers \cite{rfc4271}.
Each BGP-speaking router of an AS can decide to accept or reject incoming announcements based on the prefix itself, the AS path, or other attributes that are attached to a route (\eg BGP community values).
Due to this concept, every single AS (and, in fact, also all its routers) may have a unique viewpoint into the Internet's routing ecosystem~\cite{roughan201110}.

Many popular BGP guidelines recommend the rigorous filtering of prefixes that encompass only a few addresses~\cite{rfc7454,ripe399,ripe532,noction2020prefix,nlnog2020filtering,Gert2013transit,MANRS2021Filter} and, hence, those prefixes have been shown to propagate neither far nor reliably~\cite{strowes2017bgp}.
While the possible reasons for announcing these types of prefixes are broad and range from traffic engineering over multi-homing configurations to prefix-hijack prevention~\cite{cittadini2010evolution,huston2017bgp}, the boundary for announcements which are deemed ``widely acceptable'' are usually considered to be a /24 prefix in IPv4 and a /48 prefix in IPv6.

In this paper, we perform an in-depth analysis of prefixes that are more specific than those boundaries, \ie /25 to /32 IPv4 prefixes and /49 to /128 IPv6 prefixes.
We refer to those prefixes as \textbf{\hsps} (\hspsS, see \Cref{appendix:definition} for more details) and analyze their prominence in the global routing ecosystem, the functions that they serve, and whether they represent intentional or accidental announcements.
More specifically, we make the following main contributions:%

\noindent\textbf{Observability.}
We perform a decade long analysis of \hspsS as seen by 67 route collectors %
(see \Cref{sec:mot}).
We find that the number of \hspsS has increased substantially since 2010 and peaked in 2018 at around 115K IPv4 and 18K IPv6 prefixes.
While we observe that especially \hspsS which are announced consistently for an entire year are visible by hundreds of collector peers, the average HSP can only be seen by a handful of them.

\noindent\textbf{Use Cases \& Functions.}
We analyze potential use cases of \hspsS by combining insights from different analyses of CIDR sizes, BGP communities, and service hit rates across multiple years (see \Cref{sec:func}).
We find that IPv4 \hspsS mostly represent (internal) routes towards peering subnets and blackholing, whereas IPv6 \hspsS are mainly used for address block relocations and, in substantially fewer cases, blackholing.
We further find that \hspsS are unlikely to contain many end hosts and that they are rarely used for traffic engineering.

\noindent\textbf{Intended or Accidental Use.}
We compare the HSPs visible in BGP with those that were explicitly entered into routing databases---in particular, the Internet Routing Registries (IRR) and Resource PKI (RPKI)---to investigate intended or accidental use of \hspsS (see \Cref{sec:intent}).
We find that while thousands of ASes explicitly specify their intent to use \hspsS, many HSPs likely represent accidentally leaked routes.

\noindent\textbf{The Future of \hspsS.}
We discuss how the research and operator communities could make use of \hspsS in the future.
Finally, we publish a dashboard providing up-to-date \hspS statistics to help AS operators in detecting leaked internal routes at \url{https://hyperspecifics.io}.
\begin{figure*}[!htb]
  \minipage[t]{0.49\textwidth}
  \centering
    \includegraphics[width=\linewidth]{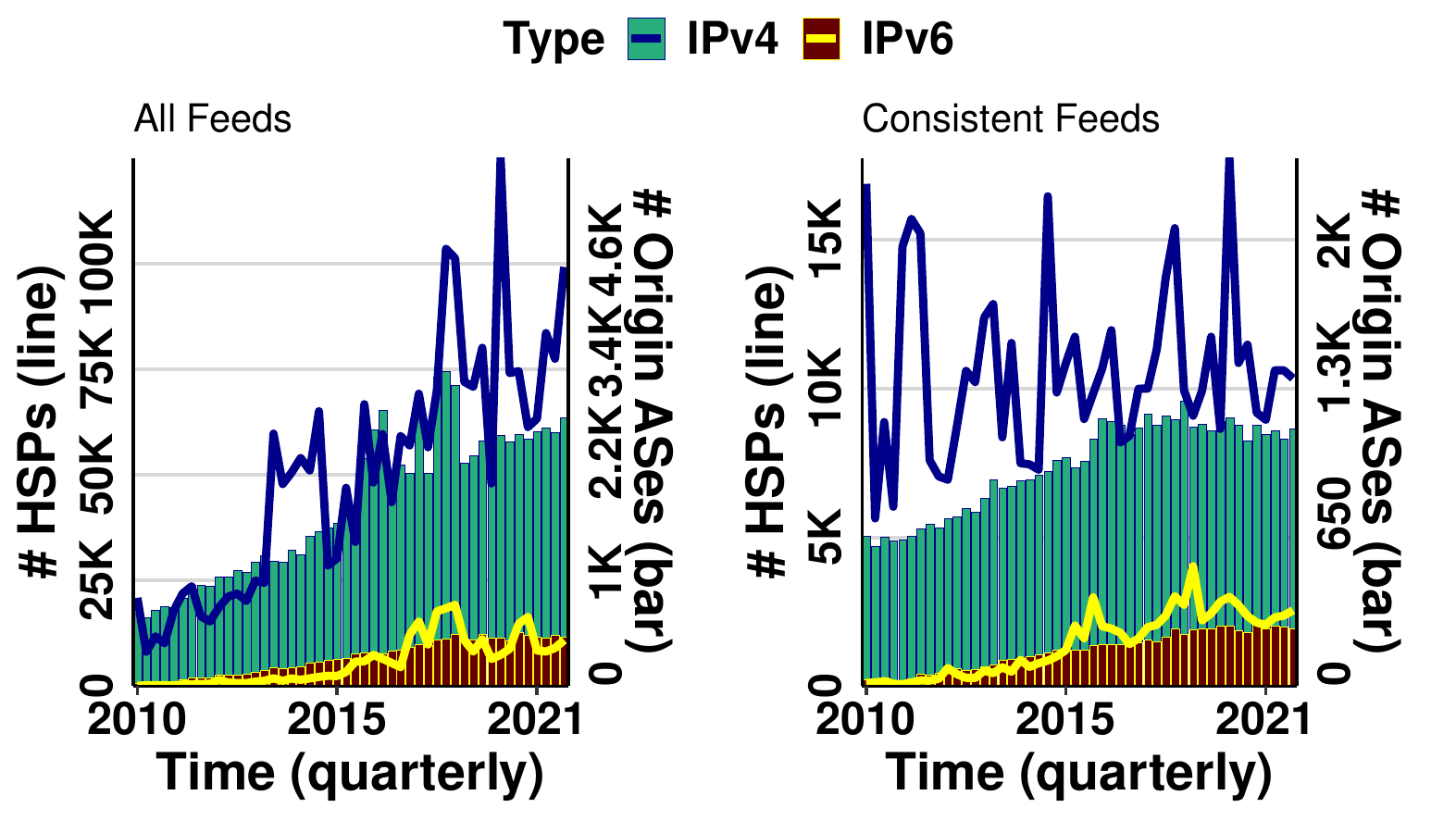}
    \caption{Growth of HSPs and HSP origin ASes as visible in all feeder ASes (left) and a consistent set of feeder ASes (right).} %
    \label{fig:mot:incr}
  \endminipage
  \hfill
  \minipage[t]{0.49\textwidth}
    \centering
    \includegraphics[width=\linewidth]{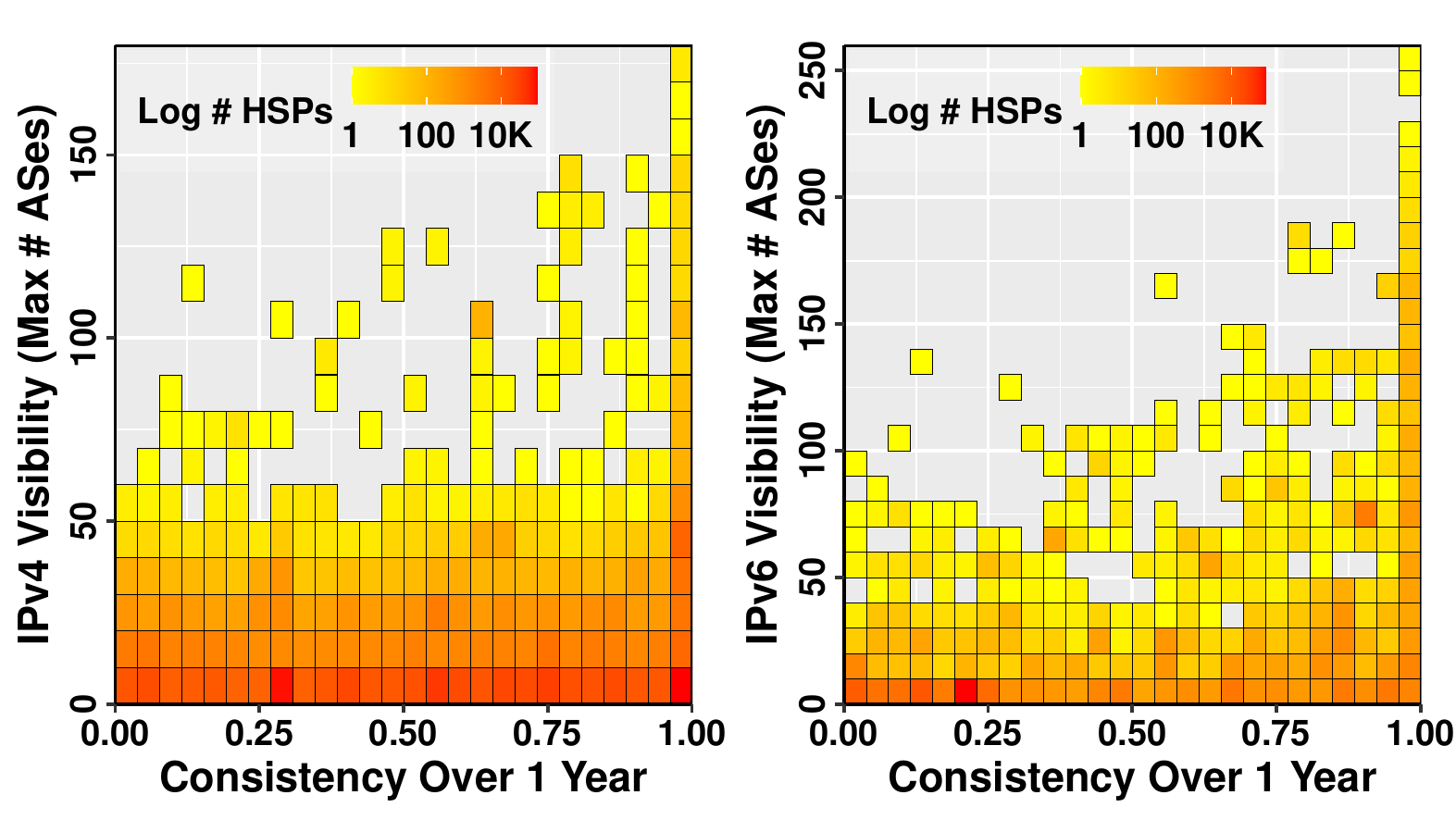}
    \caption{Heatmap showing HSP visibility and consistency for IPv4 (left) and IPv6 (right).}
    \label{fig:mot:msp_vis_vs_cons}
  \endminipage
 \end{figure*}

\section{Observability}\label{sec:mot}

We begin our exploration of hyper-specific prefixes by analyzing their current and past presence in the Internet's routing ecosystem.

In particular, we examine the routing information from hundreds of globally distributed ASes---called ``feeder ASes'' or ``route collector peers''---collected by the Isolario~\cite{isolario2021data}, RIPE RIS~\cite{ripe2021ris}, and Routeviews~\cite{routeviews2021data} projects.
Starting from January 2010, we generate snapshots consisting of a week of RIB and update files every three months until October 2021.
We provide further details about the choice of this window size in \Cref{appendix:rc_consistency}.
We employ various filtering steps to sanitize the data from, e.g., announcements of unallocated Internet resources, certain noisy origin ASes\footnote{These ASes announced either \one an extraordinary high number of HSPs (i.e., 100 or more times higher than in other snapshots) or \two HSPs in an extraordinary high number of anchor prefixes for a limited time.}, or temporarily misconfigured feeder ASes.
We also reached out to operators of noisy origin ASes.
Two of these operators were not aware of this problem, but addressed it quickly upon our notification.
A comprehensive list with justifications for the individual steps can be found in \Cref{appendix:filtering}.

First, we investigate the evolution of \hspsS from January 2010 to October 2021.
\Cref{fig:mot:incr} shows the number of hyper-specific prefixes (lines) and ASes that originate them (bars) over time. Looking at the left sub-plot, we observe that the number of seen HSPs (despite being noisy) consistently increases throughout the eleven years.
We see more than 10k IPv6 and 100k IPv4 HSPs by the end of 2021, \ie approximately one-tenth of all visible prefixes are hyper-specific (see \Cref{appendix::rca:analysis} for further details).
Relative to the increase in HSPs, we also observe an increase of ASes that originate them, with 584 and 2.5K ASes announcing hyper-specific prefixes via IPv6 and IPv4 by the end of 2021, respectively.

Given that the route collector projects acquired feeder ASes within our observation period, the increasing trend could simply be a sampling error.
To test this hypothesis, we replicate the analysis using only data from the 105 IPv4 and 45 IPv6 feeder ASes that are consistently peering with route collectors throughout all snapshots.
While our observations remain similar for IPv6, there are two changes for IPv4: \one the number of hyper-specific prefixes that can be seen by a consistent set of ASes appears more stable (if any trend exists, it remains hidden behind the massive fluctuations); and \two despite an initial increase, the number of ASes originating HSPs stagnates after 2016.
Therefore, the number of IPv4 HSPs does not show a constant increase over time, but rather we observe more IPv4 HSPs due to an increase in feeder ASes at route collector projects.

This hypothesis check leads to another observation: When shrinking the set of feeder ASes, the number of HSPs and their respective origin ASes drops substantially (note the different y-axes for the left and right subplot of \Cref{fig:mot:incr}).
To improve our understanding of this insight, we analyze the visibility of HSPs, \ie by how many peers each HSPs is seen.
At the same time, we want to understand what causes the substantial fluctuations in the number of HSPs; hence, we also analyze their consistency, \ie the fraction of time for which the prefix was seen by at least one feeder AS.
Given that a one-week observation period would not provide much insight into consistency patterns, we conduct this analysis using data from the entirety of 2020.
We first read the RIB snapshots from January 1, 2020 and then apply all updates for the whole year sequentially.
By tracking the state of each routing table on a per-update basis, we can extract consistency in seconds granularity.

\begin{figure*}[!htb]
  \minipage[t]{0.49\textwidth}
   \centering
   \includegraphics[width=\linewidth]{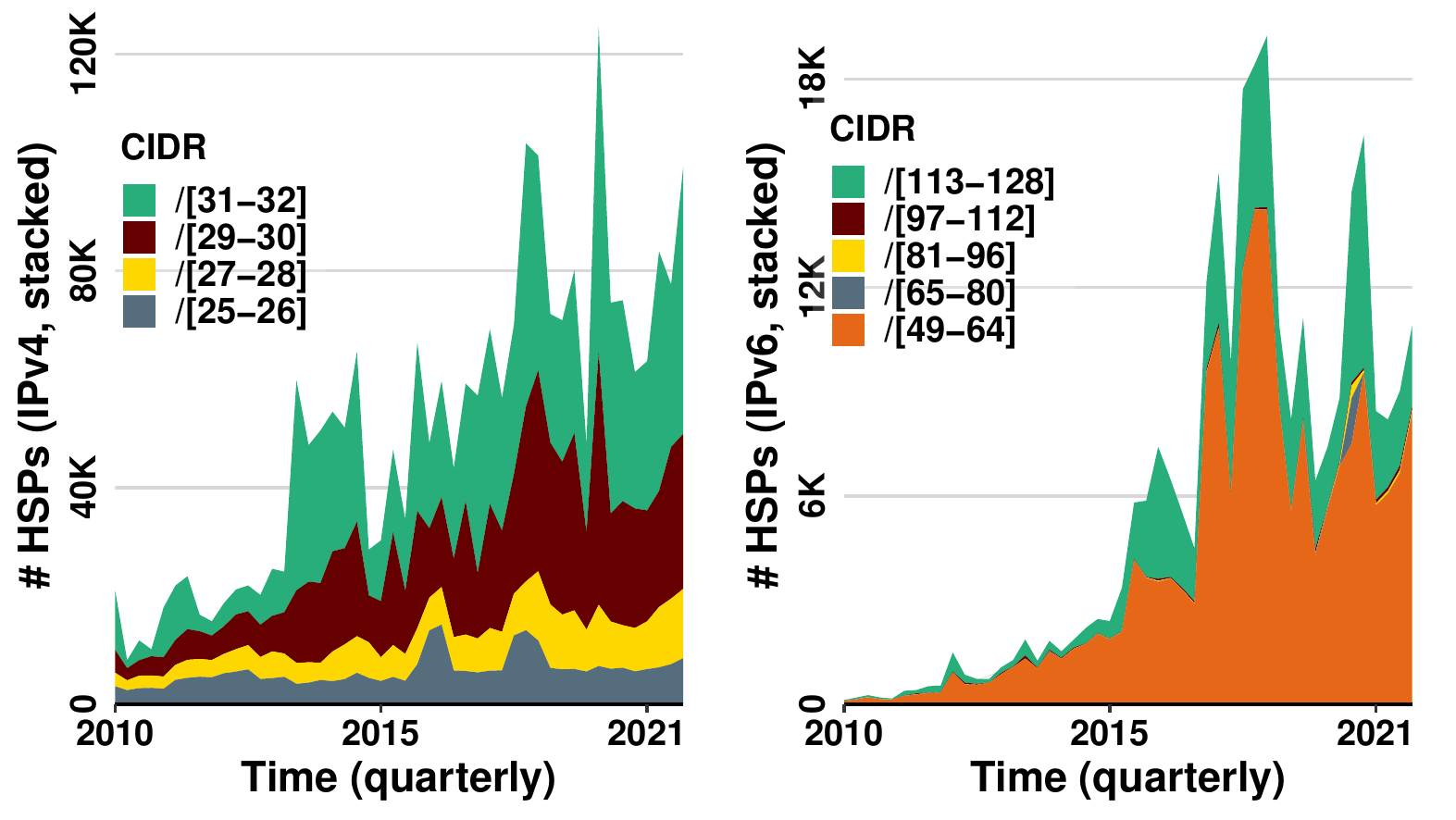}
   \caption{HSPs per CIDR size over time.}\label{fig:func:msp-by-cidr}
  \endminipage
\hfill
\minipage[t]{0.49\textwidth}
    \centering
    \includegraphics[width=\linewidth]{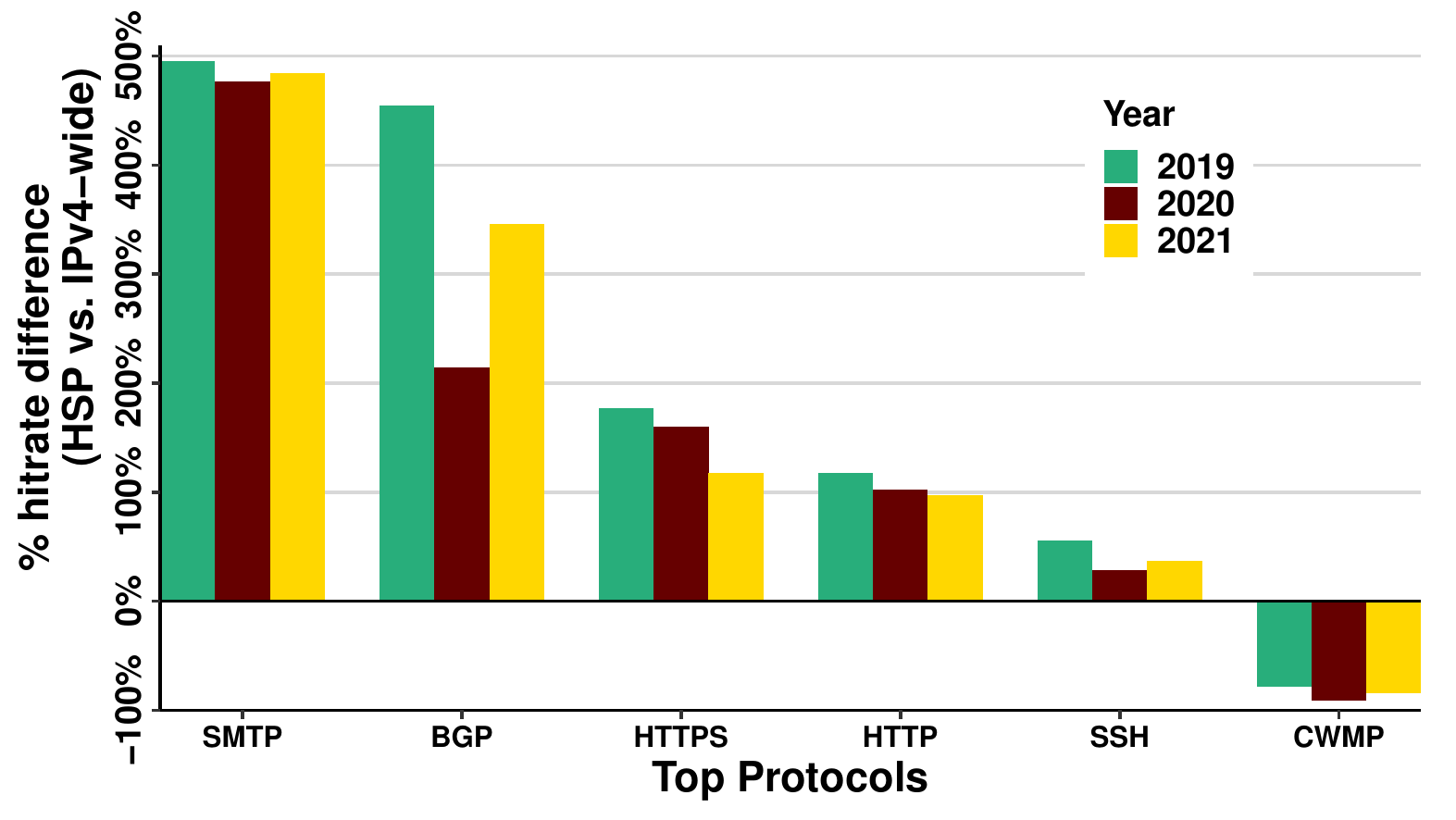}
    \caption{Hit rate comparison of HSPs vs. IPv4-wide.}\label{fig:rapid7_hsp_vs_ipv4_hitrate}
    \endminipage
\end{figure*}

\Cref{fig:mot:msp_vis_vs_cons} reports the visibility of an HSP on the y-axis against its consistency on the x-axis. For both heatmaps---IPv4 (left) and IPv6 (right)---each cell represents groups of ten feeder ASes on the y-axis and two weeks of time on the x-axis.
We first observe that there is no particular consistency trend: While some \hspsS can only be observed for less than two weeks, others can be observed throughout the entire year.
Our second observation is that the vast majority of hyper-specific prefixes can only be observed by a small number of collector peers, although we do also observe \hspsS being visible during the entire year by hundreds of peers.
This observation aligns with the restricted propagation characteristics of HSPs reported by previous blog posts~\cite{aben2014propagation,aben2015has,strowes2017bgp} and observed by our own active experiments (an in-depth description of the experiments, their analysis, and subsequent results can be found in \Cref{appendix:testbed}).
We hypothesize that the substantial fluctuations in the number of totally observed HSPs is a result of these two observations;
the restricted propagation of HSPs might inflate the importance of the individual placement of feeder ASes and HSP origin ASes, and the tens of thousand of short-lived HSPs might cluster around certain real-world events, such as DDoS attacks or data center outages.

\textbf{In summary,} we observe that the presence of hyper-specific prefixes in the Internet's routing ecosystem has increased through the last decade and \hspsS make up about one-tenth of all the prefixes that are observed by route collectors.
In IPv4 the increase in \hspsS is driven by an increment in feeder ASes, whereas in IPv6 we see an increase also for a constant set of feeder ASes.
While most HSPs only propagate locally, some of them are globally visible and can be consistently observed throughout an entire year.
\section{Use Cases \& Functions}~\label{sec:func}

Given their past and current presence in the global routing system, we want to get a deeper understanding of the functions that hyper-specific prefixes potentially serve.
As a first step in this direction, we use the fact that specific CIDR sizes often hint towards certain use cases.
Consider the following example: If an AS wants to defend one of its servers against an ongoing DDoS attack, it may use blackholing announcements.
Up to 98~\% of these announcements are /32 (/128) IPv4 (IPv6) prefixes, \ie they only cover the specific addresses of the attacked servers~\cite{giotsas2017inferring,dietzel2016blackholing,dietzel2018stellar}.
Larger CIDR sizes are rarely used for blackholing, as they would impair the services running on non-attacked servers as well, \ie they would introduce unnecessary collateral damage~\cite{nawrocki2019down}.
Using similar lines of reasoning, we rely on the following associations between CIDR sizes and intended use cases:
We associate \one /25 and /26 IPv4 prefixes with traffic engineering (\eg selective announcements\cite{quoitin2003interdomain,chang2005inbound}), \two /29 and /30 IPv4 prefixes with (Point-to-Point) peering subnets (\ie the subnets needed to form inter-AS connections)~\cite{rfc3021}, \three /31 and /32 IPv4 prefixes with blackholing~\cite{giotsas2017inferring,dietzel2016blackholing,dietzel2018stellar}, \four /49 to /64 IPv6 prefixes with address block reassignments~\cite{padmanabhan2020dynamips}, and \five /113 to /128 IPv6 prefixes again with blackholing\footnote{In private conversations a large European IXP confirmed that around 90 \% of all blackholed IPv6 prefixes fall into the /113 to /128 prefix range.}.

\Cref{fig:func:msp-by-cidr} shows the number of IPv4 (left) and IPv6 (right) HSPs over time colored by their respective CIDR size groups.
We first observe that the overall trends are stable over time.
In IPv4, we observe that the most common CIDR size is /31--/32, \ie the most prominent use case seems to be blackholing.
Yet, we also observe that /29--/30 HSPs are comparably common; hence, many HSPs may actually represent peering subnets.
Given that only about 10~\% of HSPs have a CIDR size of /25 or /26, we believe that traffic engineering is a rare use case.
For IPv6, we mainly observe the /49--/64 CIDR size range that we associate with address block relocations.
In some ASes we also observe instances of /64s being used by hypergiants for off-nets \cite{gigis2021seven}.
We further observe a small fraction of /113--/128 CIDR sizes that we associate with blackholing.
The share of blackholing \hspsS is smaller in IPv6 compared to IPv4, which is in line with reports that blackholing in IPv6 makes up less than 2 \% compared to IPv4 \cite{nawrocki2019down,giotsas2017inferring}.
Those observations also explain some of the fluctuations that we observed in the previous section---blackholing events, and their subsequently announced prefixes, are often short-lived~\cite{nawrocki2019down} and subsequently can cause substantial changes in the number of unique HSPs seen throughout a week.

As our CIDR-based analysis only provides us with hints on the actual usage, we now also analyze the services hosted in hyper-specific prefixes.
For this analysis, we leverage archived scanning data from Rapid7's Open Data platform \cite{rapid7} for 2019, 2020, and 2021.
Rapid7 frequently scans the entire routed IPv4 address space\footnote{Except for prefixes on their blocklist which were explicitly requested by network operators.} for more than 100 well-known TCP and UDP ports.
To compare regular with hyper-specific prefixes, we rely on the difference in protocol hit rate, i.e., we compare the fraction of responding hosts and total tested hosts\footnote{Given that Rapid7 does not publish the state of their blocklist, we assume that all (at the time of the scan) routed IP addresses were tested. Additionally, we focus on analyzing what services are prominent in \hspsS. We can not ensure that Rapid7 (or its upstream) does in fact receive the \hspS announcements, as information about their probing vantage points and routing is not available.} on a per-protocol basis.
We observe that four out of the top five protocols with the highest hit rate for regular and HSP prefixes overlap; BGP is only present in the HSP top five while CWMP is only present in the IPv4-wide top five.
For those six protocols, \Cref{fig:rapid7_hsp_vs_ipv4_hitrate} shows a the relative difference of hit rates between regular and hyper-specific prefixes, where a positive value indicates an increase of hit rate in hyper-specific prefixes.
While HTTP and HTTPS overall only see an increase of +100~\%, we observe strong differences when drilling down on a per-CIDR level:
When considering only /32 prefixes, HTTP's hit rate increases by more than +500~\% compared to its hit rate for IPv4-wide scans---which substantiates the association of the /32 CIDR size for blackholing.
Even more pronounced than HTTP(S), SMTP and BGP see increases of up to +500\%.
When digging deeper we further observe that BGP is mainly prevalent in /30 and /29 prefixes, which underlines that these sizes might be dedicated to routing infrastructure.
In contrast, we observe the only hit rate decrease (of more than 90\%) for CWMP---a protocol used to remotely manage customer-premises equipment (CPE) devices such as home routers~\cite{cwmp}.

\begin{figure}[t!]
 \centering
 \includegraphics[width=.8\linewidth]{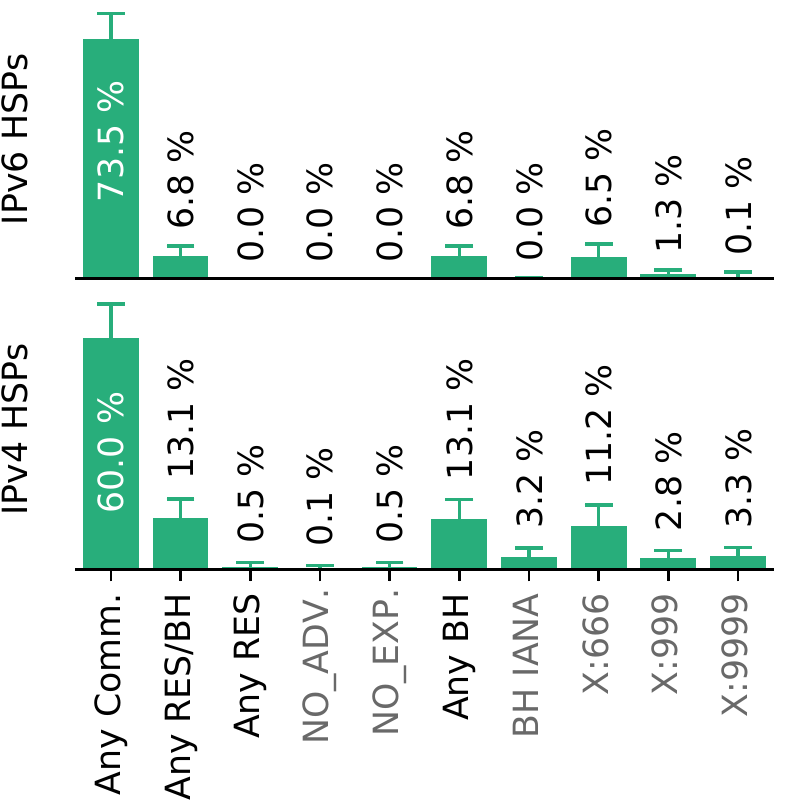}
 \caption{BGP communities distribution for \hspsS.}\label{fig:int:bgp-comms}
\end{figure}

Finally, we investigate BGP communities attached to \hspS announcements.
BGP communities are used for many different reasons, such as information tagging, blackholing, or route redistribution.
The most common BGP communities attached to \hsps are route steering or prepending instructions.
In our analysis we look for BGP communities which are specifically used for blackholing (BH) \cite{rfc7999} or restrict route propagation (RES)\footnote{We also test for communities such as NOPEER or NO\_EXPORT\_SUBCONFED, but these are not prevalent among \hspsS.}.
\Cref{fig:int:bgp-comms} shows the use of BGP communities among \hspsS from snapshots between 2019 and 2021.
The bars indicate the median share of \hspsS with the respective community, the whiskers denote the standard deviation over time. The ``Any'' keyword is used to specify groups of community targets, e.g., ``Any RES'' describes all prefixes that have any restriction community attached (i.e., it refers to the union of prefixes with ``NO\_ADV'' community and prefixes with ``NO\_EXP'' communities); similarly, the ``Any Comm.'' bar refers to the highest aggregation, i.e., prefixes for which we saw any community attached.  
As we can see, 60\% of all IPv4 \hspsS and almost three quarters of IPv6 \hspsS come with some form of BGP communities.
The vast majority of these communities is, however, not related to blackholing or restricting propagation.
Only about 13\% and 7\% of prefixes can be associated with blackholing for IPv4 and IPv6, respectively.
The by far most popular blackholing community is \texttt{X:666}.
Moreover, we see no propagation restriction communities (``no advertise'' or ``no export'') in IPv6 and only about 0.5\% in IPv4.
Furthermore, we see that RES communities are a subset of BH communities, hinting that operators do not want their blackholing prefixes to propagate.
Blackholing is therefore one contributor of \hspsS, but blackholing communities are not present on the majority of \hspS announcements.
We note that the blackholing communities that we see at route collector peers is a lower bound:
Blackholing communities---similar to other communities---could be cleaned along the path but the prefix itself could continue to propagate \cite{krenc2021level}.

\begin{figure*}[!t]
\minipage[t]{0.49\textwidth}
  \centering
  \includegraphics[width=\linewidth]{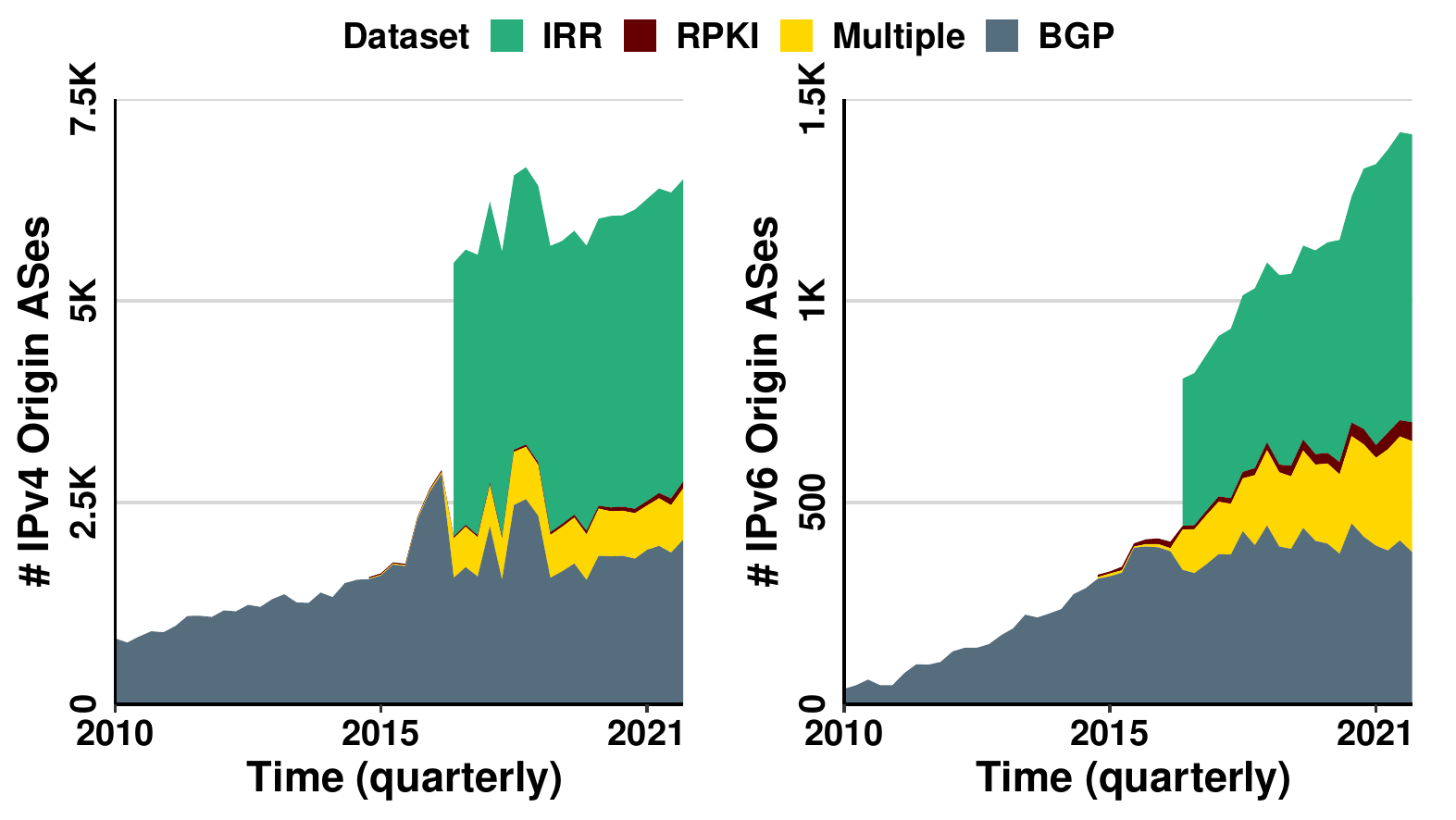}
  \caption{Visibility of origin ASes across data sets.}\label{fig:int:origins_p_datasets}
\endminipage
\hfill
\minipage[t]{0.49\textwidth}
 \centering
 \includegraphics[width=\linewidth]{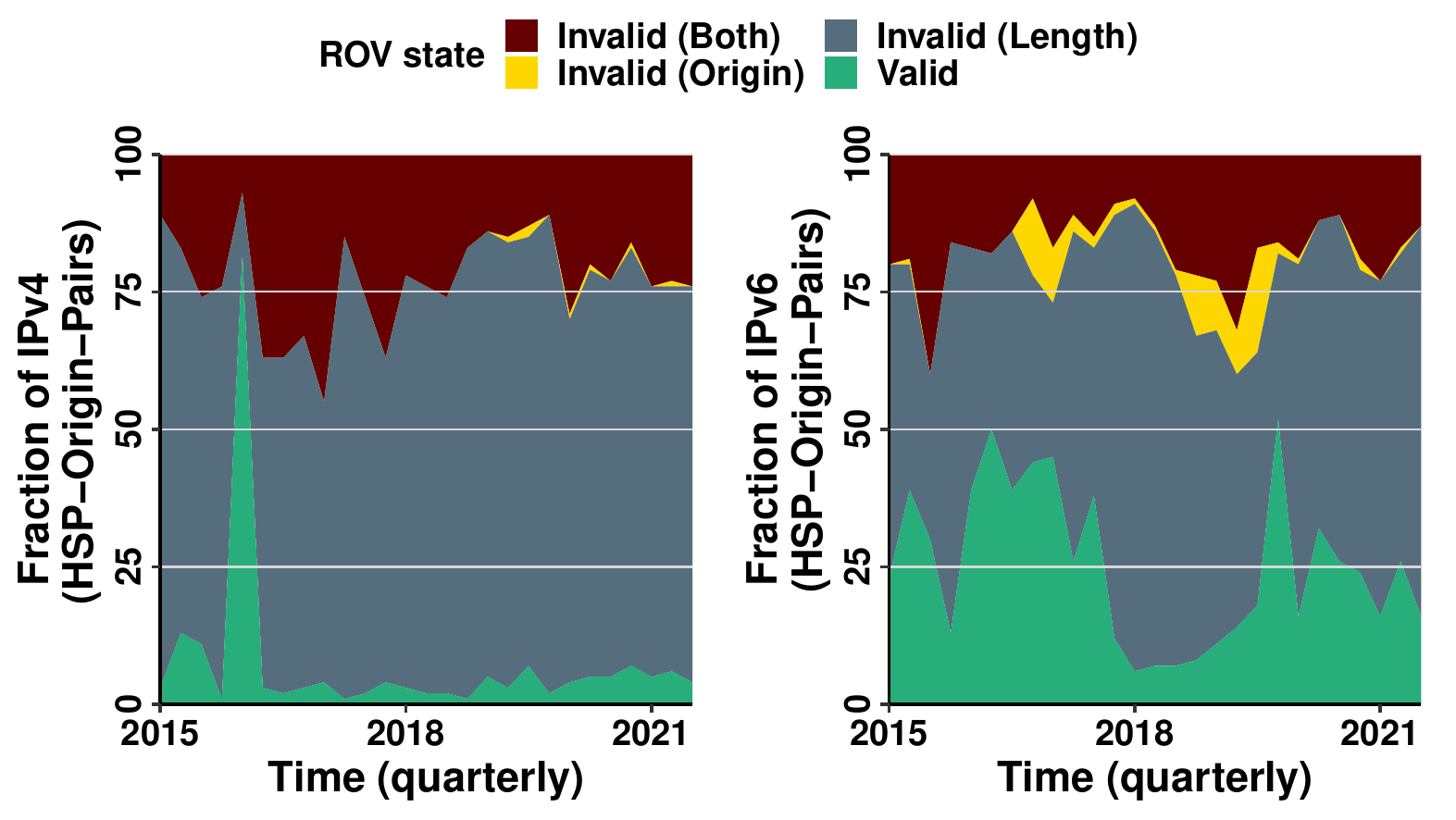}
 \caption{ROV status for HSPs}\label{fig:int:hsp-rov-status}
\endminipage
\end{figure*}

\textbf{In summary}, we observe that for IPv4 many and for IPv6 some HSPs are likely related to blackholing activities due to the used HSP prefix size.
We find concrete evidence for 7--13\% of HSPs explicitly tagged with blackholing communities.
While we also observe many HSPs dedicated to routing infrastructure (e.g., peering subnets or address relocations), we observe that hyper-specific prefixes rarely contain any CPE devices.
\section{Intended or accidental Use?}\label{sec:intent}

Now that we have a basic understanding of the use cases of \hspsS, we want to analyze whether HSPs are used intentionally or accidentally by ASes and their operators. If operators take the time and effort to explicitly enter hyper-specific prefixes into voluntarily-maintained databases, then it is likely that they plan to use them. Hence, we look at the Resource Public Key Infrastructure (RPKI) and Internet Routing Registry (IRR) operator databases.

We use private, three-monthly IRR snapshots \cite{irr2021dumps} between January 1, 2017, and October 7, 2021, which contain information about routing policies.
The RPKI database contains legally binding mappings between Internet resources and ASes. We use daily snapshots of the RPKI database \cite{rpki} from April 1, 2015, until October 7, 2021, generated by Chung \etal \cite{chung2019rpki} to verify the validity of HSP announcements by ASes.

While we extract HSPs directly from the \texttt{route(6)} objects contained in the IRR databases, the Route Origin Authorization (ROA) objects in the RPKI snapshots describe CIDR size ranges~\cite{rfc6483}.
Hence, a ROA can explicitly describe an HSP when both the minimum and maximum prefix length are hyper-specific, or implicitly when only the maximum prefix length is hyper-specific.
When extracting HSPs and their origins from the RPKI database, we rely solely on explicit definitions as these clearly represent the desire to use HSPs (as all covered prefixes are hyper-specific). As implicit definitions might describe the future---but not necessarily current---use of HSPs (e.g., an AS might currently announce a /24 but has already entered a currently unused max-length of /25), we decide to ignore them. 
We compare the HSPs on those two databases against the HSPs visible via BGP route collectors.

\Cref{fig:int:origins_p_datasets} shows the number of unique origin ASes for both IPv4 and IPv6 within each dataset over time.
We classify those origin ASes available in more than one dataset into the ``Multiple'' category.
Our first observation is that for both IPv4 and IPv6, the IRR dataset contains the largest fraction of HSP origin ASes.
While this might imply that network operators tend to actually use HSPs, it is well-known that route objects can become stale given that the database is only maintained on a voluntary basis~\cite{siganos2004analyzing}.
Yet, some entities, e.g. certain IXP Route Servers~\cite{decix2021irr}, require route objects in the IRR database to redistribute prefixes (i.e., HSPs).
Even for the RPKI database we observe hundreds of explicitly defined HSPs\footnote{Most of these HSPs are also in the BGP data set and hence end up up in the multiple class.}.
Notably, for the last snapshot in October 2021, implicit HSPs would have increased the number of RPKI origin ASes from 294 to 990 for IPv4 and from 172 to 794 for IPv6, respectively.
Beyond these intentional HSPs, we also observe that many of the HSPs from Route Collectors have no entries in operator databases, hence, they could potentially represent accidental announcements or misconfigured route collector sessions that leak internal routes.

While it is hard to link malicious intent to a more-specific announcement (since it could be, e.g., an address leasing agreement~\cite{prehn2020wells} or traffic engineering of sibling ASes \cite{gao2001inferring}), we want to understand if the visible HSPs in the BGP are legitimate prefix advertisements by valid origin ASes or associated with possible prefix hijacks. Therefore, we perform route origin validation (ROV) of HSPs and its origin AS by checking them against the ROA records from the RPKI dataset. If a ROA covers the address space described by the prefix, then this prefix can violate the ROA in two ways: it can be too specific---which we mark as ``Invalid (Length)''---and it can be announced by a different origin---which we mark as ``Invalid (Origin).'' If both of these conditions are met at the same time, we mark a prefix as ``Invalid (Both).'' If none of these conditions are met, we consider the
prefix as ``Valid.''
Notably, we observe that 22~\% of IPv4 and 19~\% of IPv6 HSPs have a covering ROA entry (median percentages across snapshots in 2020 and 2021). 

\Cref{fig:int:hsp-rov-status} shows that legitimate ASes, \ie the valid and invalid length categories together, advertise around 75~\% of all HSPs.
With an average of 25~\% and peaking to around 50~\% in 2016, 2017, and 2019, IPv6 has a higher percentage of valid HSPs than IPv4. The HSPs with invalid length form the largest group in IPv4, and mostly the second largest group in IPv6.
The third largest group of HSPs has the ``Invalid (Both)'' ROV state, while the invalid origin category forms a minor fraction of HSPs' ROV state. Legitimate ASes advertise around 75~\% of HSPs, which indicates that HSPs are not majorly associated with BGP prefix hijacks.
Beyond malicious ASes, the ``Invalid (Origin)'' and ``Invalid (Both)'' status could also be caused by not properly entered sibling ASes \cite{gao2001inferring} or from a DDoS Protection Service (DPS) \cite{jin2018your}.
We analyze how many \hsps are originated from a DPS as identified by Jin \etal \cite{jin2018your} and find that only around 1~\% of \hspsS in IPv4 and IPv6 are related to DPS companies.

\textbf{In summary,} we observe that for both IPv4 and IPv6, hundreds of ASes intentionally entered hyper-specific prefixes into operator databases.
Yet we also see that many of the HSPs that are visible from route collectors have no respective entries and are likely related to the accidental announcement or disclosure of internal routes.
This is further substantiated by the observation that most HSPs are actually ROV invalid since they are more specific then intended by their covering ROA entry.

\section{Discussion}
\label{sec:discussion}

\noindent\textbf{Research Community.} 
While many \hspsS seem to be intentional, we also observe a large number that potentially represent leaked internal routes. 
While the task of reconfiguring a leaking router ultimately belongs to the feeder AS' operators, we believe that the maintainers of route collector projects
play a vital role when it comes to raising awareness for the existing problems. 
To support and guide this process, we publish and maintain a dashboard that provides up-to-date \hspS statistics as well as a rankings of the top HSP contributors at \url{https://hyperspecifics.io}.
Beyond fixing potential leakage errors, we believe that studying the potential correlations between hyper-specific prefixes and their less-specific counter parts may lead to new insights into the routing optimizations used by ASes.

\noindent\textbf{Operator Community.} Even though various guides~\cite{noction2020prefix,rfc7454,nlnog2020filtering,Gert2013transit,MANRS2021Filter} recommend strict filtering of \hspsS, we find that many hyper-specific prefixes propagate to 100 or more collector peers. 
After discussing our results with thirteen operators from different types of networks, we believe that the limited filtering is often a result of popular customer requests. 
The operator of a major transit network told us that their network recently (throughout Summer 2020) changed from the filtering of all IPv4 HSPs to only filtering prefixes more specific than /28; this shift enabled (especially new and small) customer networks to perform basic traffic engineering despite a limited address allocation\footnote{This is a direct result of the current IPv4 Address exhaustion and the subsequently inflated prices~\cite{prehn2020wells}.}. 

This opens up the question whether operators should filter HSPs in the first place. 
We believe that for IPv6 the answer is a resounding ``yes''. 
Given that there is no shortage of IPv6 addresses and obtaining new blocks is virtually free (compared to the high costs of obtaining IPv4 addresses), we do not see any reason to loosen the current filtering guidelines.
For IPv4, we think that the answer should be more nuanced. While loosening the filtering guidelines allows even small ASes to perform traffic engineering, it would also further increase the routing table size. 
Hence, we believe that shifting the acceptable boundaries by a few CIDR sizes (e.g., /26 or /28) might be an agreeable compromise.

\section{Related Work}
\label{sec:related-work}

In this section, we report on related work in the areas of \hsp analysis and prefix deaggregation.

\noindent\textbf{\hspS Analysis:}
Previous research in this area consists mostly of blog posts. %
In 2014, Aben and Petrie report on an experiment where they announced /24, /25, and /28 IPv4 prefixes and ran \ra measurements to them~\cite{aben2014propagation}.
Their findings show that \hspsS are visible for at most 20\perc of RIPE RIS peers~\cite{ripe2021ris} with route objects slightly improving the visibility.
The \ra experiments lead to similar results with fewer than 15\perc of probes reaching their targets.
One year later, Aben and Petrie revisit the propagation of \hsps and find a marginal increase of a few percent~\cite{aben2015has}.
In 2017, Strowes and Petrie conclude that not much has changed regarding \hsp propagation and at most one fourth of all BGP peers receive those announcements~\cite{strowes2017bgp}.

\noindent\textbf{Prefix Deaggregation:}
In 2002, Bu \etal first characterize prefix deaggregations and the reasons for them, \eg traffic engineering, multi-homing, and address fragmentation~\cite{bu2002characterizing}.
Meng \etal report in 2005 that even newly assigned address space is deaggregated and that the deaggregation rate of prefixes increases over time~\cite{meng2005ipv4}.
In 2010, Cittadini \etal \cite{cittadini2010evolution} report that more than 10~\% of ASes deaggregate their prefixes while around 1~\% of ASes announce more than 10 prefixes for each address block they got assigned.
Lutu \etal present a simulation model that estimates that origin ASes can reduce their transit cost by 5~\% by using more-specific announcements \cite{lutu2012economic,lutu2013aftermath,lutu2015analysis}.
Notably, the authors neither focused on IPv6 nor on \hsps.
In 2016, Krenc and Feldmann analyze the address delegations realized via prefix deaggregations and report on delegations from customers to providers or between unrelated ASes (often involving CDNs)~\cite{krenc2016bgp}.
In 2017, Huston analyzes the prevalence and different types of \msp announcements in the Internet as an effect of prefix deaggregation \cite{huston2017bgp}.
His taxonomy attributes \mspsS to three different root causes, hole punching (different origin AS), traffic engineering (same origin AS, but different AS path), and overlay (same AS path).
He concludes that the former two play a useful role for network operators, while the usefulness of overlay \msps could be argued about.
Huston did not specifically investigate the effect of \hsps.

To the best of our knowledge, this paper presents the first scientific analysis of \hsps by providing an in-depth look into the prevalence and possible root causes for \hspsS in the wild. 
\section{Conclusion}
\label{sec:conclusion}

In this paper, we analyzed the presence of hyper-specific prefixes in the Internet's ecosystem throughout the last decade.
While we found an overall increase in the number of HSPs, most of them can only be observed by a few route collector peers. 
Yet, there are still plenty of HSPs that propagate to hundreds of route collector peers and can be consistently observed throughout an entire year.
Inspired by those findings, we took a closer look at the function that these prefixes serve.
For IPv4, we observed that HSPs are mainly associated with blackholing and infrastructure announcements (e.g., routes to peering subnets).
While we only found limited evidence for any connection to traffic engineering, we observed that hyper-specific prefixes are less likely to contain end-user devices.
For IPv6, we observe that almost all \hsps are related to address block reassignments, with only a small fraction representing blackholing.
Even though we have seen that hundreds of networks use HSPs intentionally, we attributed even more cases to the accidental ``leakage'' of internal routes.
Finally, we discussed the current state of HSPs from an academic as well as an operator point of view.

\balance

\bibliographystyle{ACM-Reference-Format}
\bibliography{paper}

\appendix

\clearpage

\section{MSPs vs. HSPs}
\label{appendix:definition}

In this section, we want to briefly contrast the definitions of More-Specific Prefixes (MSPs) and Hyper-Specific Prefixes (HSPs). $P$ is an MSP of $P^\prime$ when the address space that $P$ describes is entirely contained in $P^\prime$, e.g., 1.0.0.0/24 is an MSP of 1.0.0.0/22. In contrast, we call a prefix hyper-specific if its CIDR size is larger than /24 or /48 for IPv4 and IPv6, respectively. While labelling a prefix as an MSP requires another (covering) prefix, the HSP label relies entirely on the CIDR size of a given prefix and does not require a second, related prefix. Notably, many---but not all---hyper-specific prefixes are also MSPs of less-specific prefixes. As the definitions of MSPs and HSPs are very different, further classifications of HSPs (as in, e.g., Geoff Huston's blogpost~\cite{huston2017bgp}) are not directly applicable to HSPs. 

\section{Route Collector Consistency}
\label{appendix:rc_consistency}

In order to analyze representative route collector snapshots of the three RC projects Isolario~\cite{isolario2021data}, RIPE RIS~\cite{ripe2021ris}, and Routeviews~\cite{routeviews2021data}, we first analyze their consistency over time.
To estimate the consistency, we initially retrieve data for all days in 2010, 2013, 2016, and 2020.
For each day, we download the first routing information base (RIB) snapshot as well as all available update messages produced by each RC. If an update file is missing, we, additionally, download the first available RIB snapshot after the missing update file.
After extracting the \hspsS for each day, we analyze consistency as the fraction of \hspsS seen at day $n+w+1$ that are also visible within the observation period $[n, n+w]$.
Notably, we try all possible window size positions, i.e., $n \in \{0, ..., d-w-1\}$ where $d$ is the number of days in the given year.

\begin{figure}[h!]
	\centering
    \includegraphics[width=0.66\linewidth]{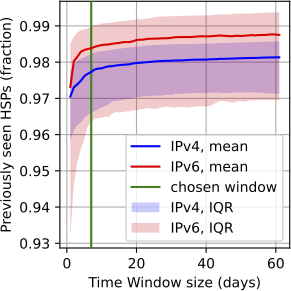}
  	\caption{Impact of window size on visible HSPs.}
  	\label{fig:datasets:cons}
\end{figure}

\Cref{fig:datasets:cons} shows the mean as well as the interquartile range (IQR) across all possible $n$ for window size $w$ between 1 and 60 days for IPv4 and IPv6 HSPs in 2020.
We observe that a seven-day window allows us to achieve a consistency of 97~\% and 98~\% for IPv4 and IPv6, respectively.
Notably, further expanding the window size to 60 days would only increase the consistency by \textasciitilde0.5\perc.
Given that we now have a snapshot aggregation window, we still need to pick a snapshot interval.
When comparing the number of visible HSPs for different snapshot intervals, we observe that a three-month interval provides an optimal balance: While the number of data points is still capable of capturing all visible trends in more-frequent snapshot intervals, the reduced amount of data (i.e., only seven days every three months) still allows us to perform computationally expensive observations for the entire decade promptly.

\section{Further analysis}
\label{appendix::rca:analysis}
\begin{figure*}[!htb]
	\minipage[t]{0.49\textwidth}
	\includegraphics[width=\linewidth]{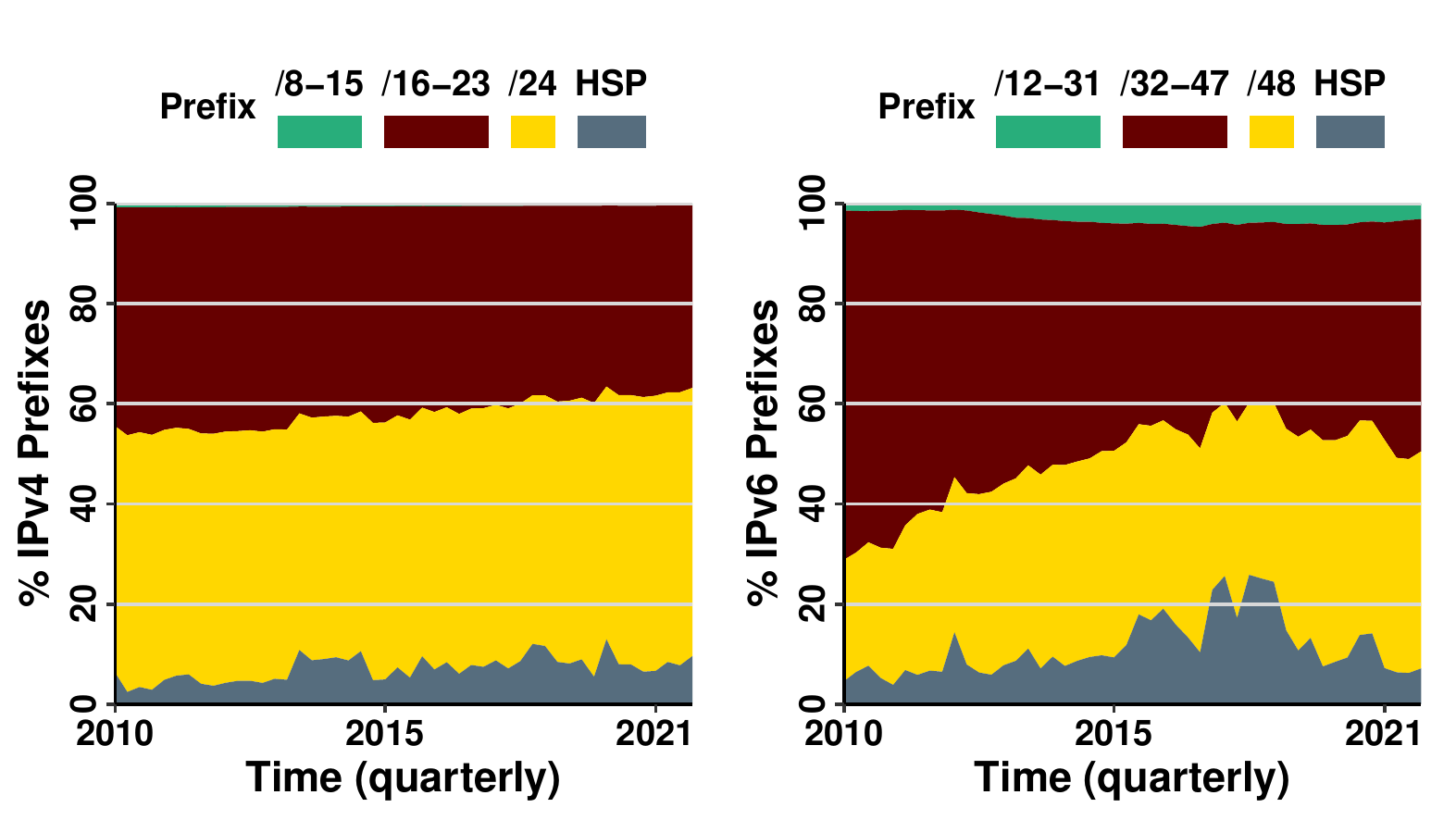}
	\caption{HSP prefix contribution over time
	}\label{fig:rca:hsp-pfx-contr}
  \endminipage
	\hfill
	\minipage[t]{0.48\textwidth}
	\includegraphics[width=\linewidth]{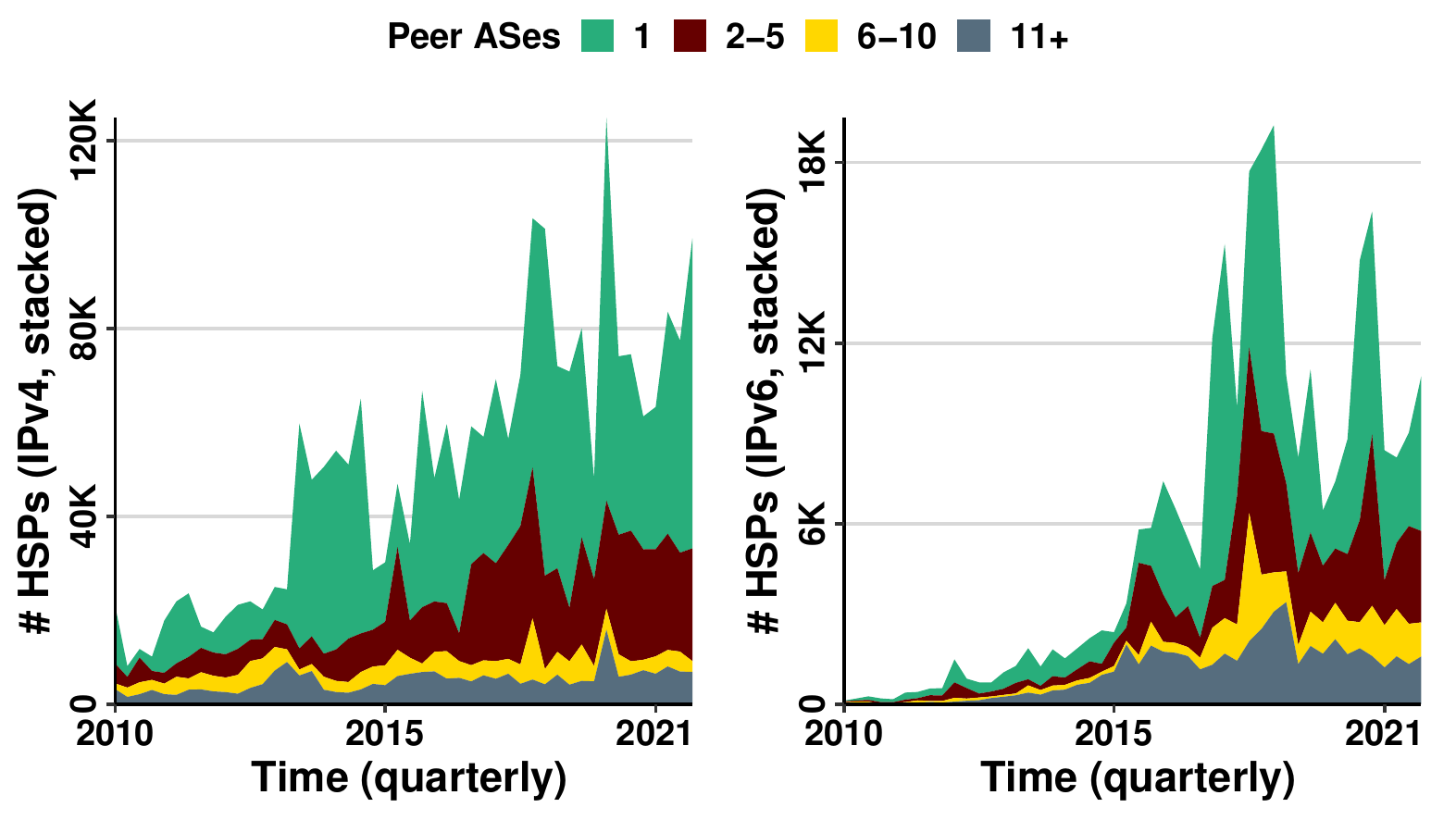}
  \caption{HSPs by \# Peer ASes over time}\label{fig:rca:msp-by-rcp}
	\endminipage
\end{figure*}

\textbf{How prominent are HSPs?} To understand the prevalence of hyper-specific prefixes, we aggregate the routing tables of all collector peers and compare the distribution of prefixes depending on CIDR sizes. Figure \ref{fig:rca:hsp-pfx-contr} shows those distributions as stacked bar plots for each snapshot. %
We observe that up to 13~\% (in 2015) and 25~\% (in 2018) of totally visible prefixes are hyper-specific for IPv4 and IPv6, respectively. Yet, the usual contribution of HSPs is approximately 10~\% for most months. Note that this does not mean that any single routing table contains that many HSPs on its own.

\textbf{How visible are HSP?} To further elaborate on this point, Figure \ref{fig:rca:msp-by-rcp} shows the number of hyper-specific prefixes per IPv4 (left) and IPv6 (right) snapshot separated based on the number of route collector peers that can see them. For IPv6, we observe that most hyper-specific prefixes can be seen by two or more peers, with around a fifth of all HSPs being visible by 11+ peers for most snapshots. Similar to the previous plot, we again observe a peak of (\textasciitilde20K) hyper-specific prefixes at around 2018. While we are not able to account this peak to a single factor, we observe that the increase is rather uniform across collector peers, origin ASes, intermediate ASes, and address space and, hence, is unlikely to stem from a measurement artifact or some local misconfiguration. When comparing the situation before and after the peak, we still can see an increase from \textasciitilde7K HSPs in 2016 to \textasciitilde11K HSPs in 2021. In contrast to IPv6, many HSPs in IPv4 can only be seen by one peer. While we observe few HSPs that can be seen by 100+ peers, the vast majority of HSPs can only be seen by 10 or less peers. Even though the number of low-visibility HSPs strongly fluctuates between snapshots, it increases rather continuously across many snapshots. Both such characteristics are significantly less pronounced for IPv4 HSPs that can be seen by 6+ peers. This difference may be accountable to various reasons including the association of a prefix to a certain function or a prefix's lifetime.

\textbf{HSP aggregation.} ASes often have economical incentives to keep their BGP routing table size low. To realize this goal, some ASes aggregate (multiple) more-specific routes into a single less-or-equally-specific route \cite{cittadini2010evolution}. If an anchor-prefix results from aggregating prefixes with different CIDR sizes (prefix-based aggregation), we know that one of such pre-aggregation prefixes must have been hyper-specific. Yet, confidently identifying such aggregations is challenging. According to RFC 4271~\cite{rfc4271}, a router \textit{MAY} set the \texttt{AGGREGATOR} field when it performs prefix-aggregation---which can serve as indication that \textit{some} form of aggregation must have happened. Thus, we first extract all routes for anchor-prefixes which have the \texttt{AGGREGATOR} field set. At this stage, our selected routes might be a result prefix-based aggregation or the aggregation of different routes---e.g., with different \texttt{AS\_PATH} attributes (path-based aggregation)---for the same prefix (or both). To reduce the likelihood of falsely identifying HSP usage due to path-based aggregation, we rely on the \texttt{ATOMIC\_AGGREGATE} field as well as the presence of \texttt{AS\_SET} elements in the \texttt{AS\_PATH} attribute. A router \textit{SHOULD} set the \texttt{ATOMIC\_AGGREGATE} field if the newly generated \texttt{AS\_PATH} attribute of the post-aggregation route does not contain all AS numbers present in the pre-aggregation routes, e.g., the paths $A B$ and $A C$ can be aggregated to $A B$ (which hides the existence of $C$). If the \texttt{ATOMIC\_AGGREGATE} field is not set, ASes often use \texttt{AS\_SET}s to signal path-aggregation, e.g., the paths $A B$ and $A C$ can be aggregated to $A \{B,C\}$ (where \{...\} denotes the \texttt{AS\_SET} containing all ASes after $A$). As the \texttt{ATOMIC\_AGGREGATE} field and \texttt{AS\_SET}s indicate path-based aggregation, we remove all anchor-routes that contain at least one of them.

\begin{figure*}[!htb]
\minipage[t]{0.49\textwidth}
  \includegraphics[width=\linewidth]{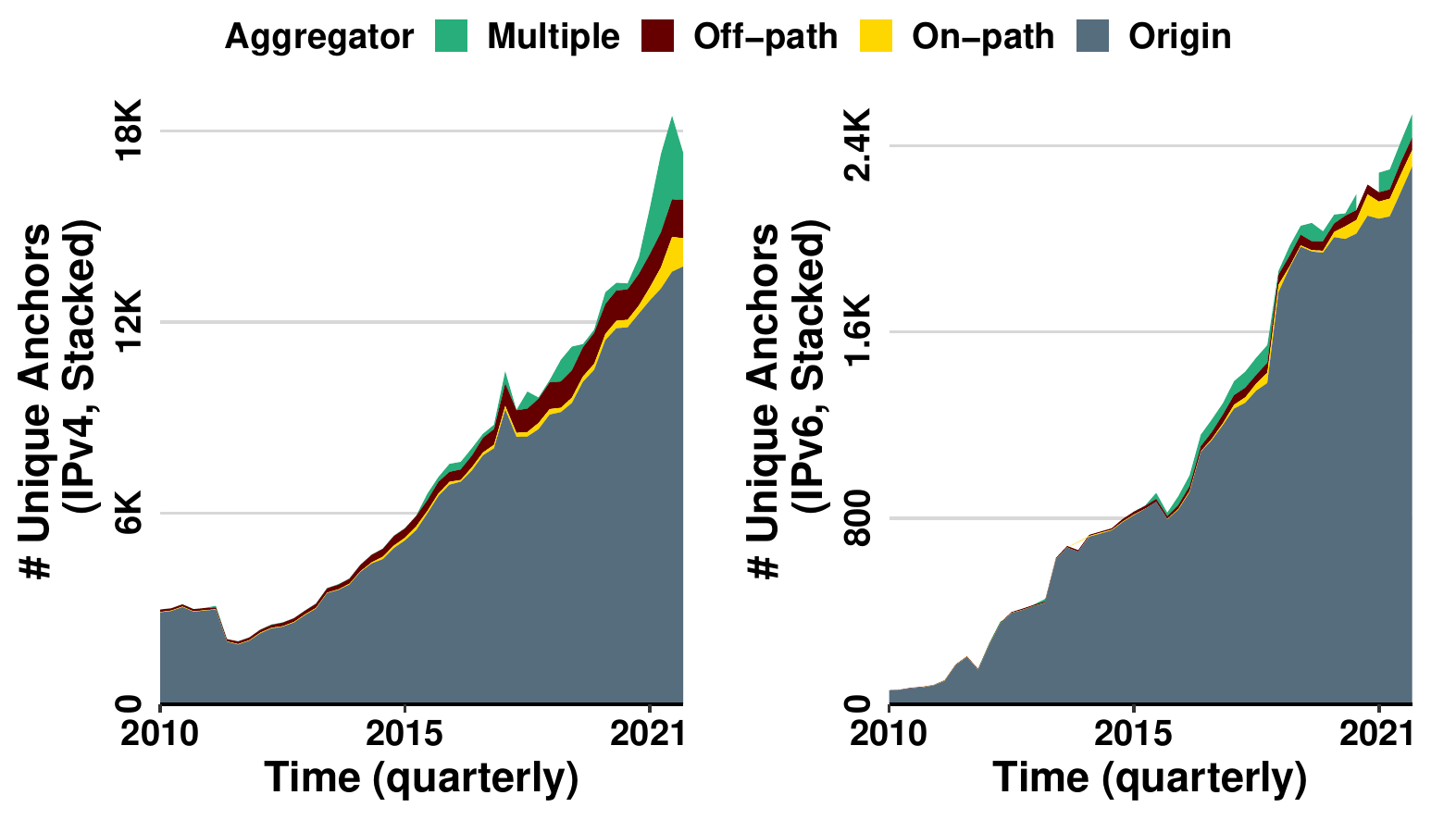}
  \caption{Position of HSP Aggregation
  }\label{fig:rca:hsp-aggr-position}
\endminipage
\hfill
\minipage[t]{0.49\textwidth}
  \includegraphics[width=\linewidth]{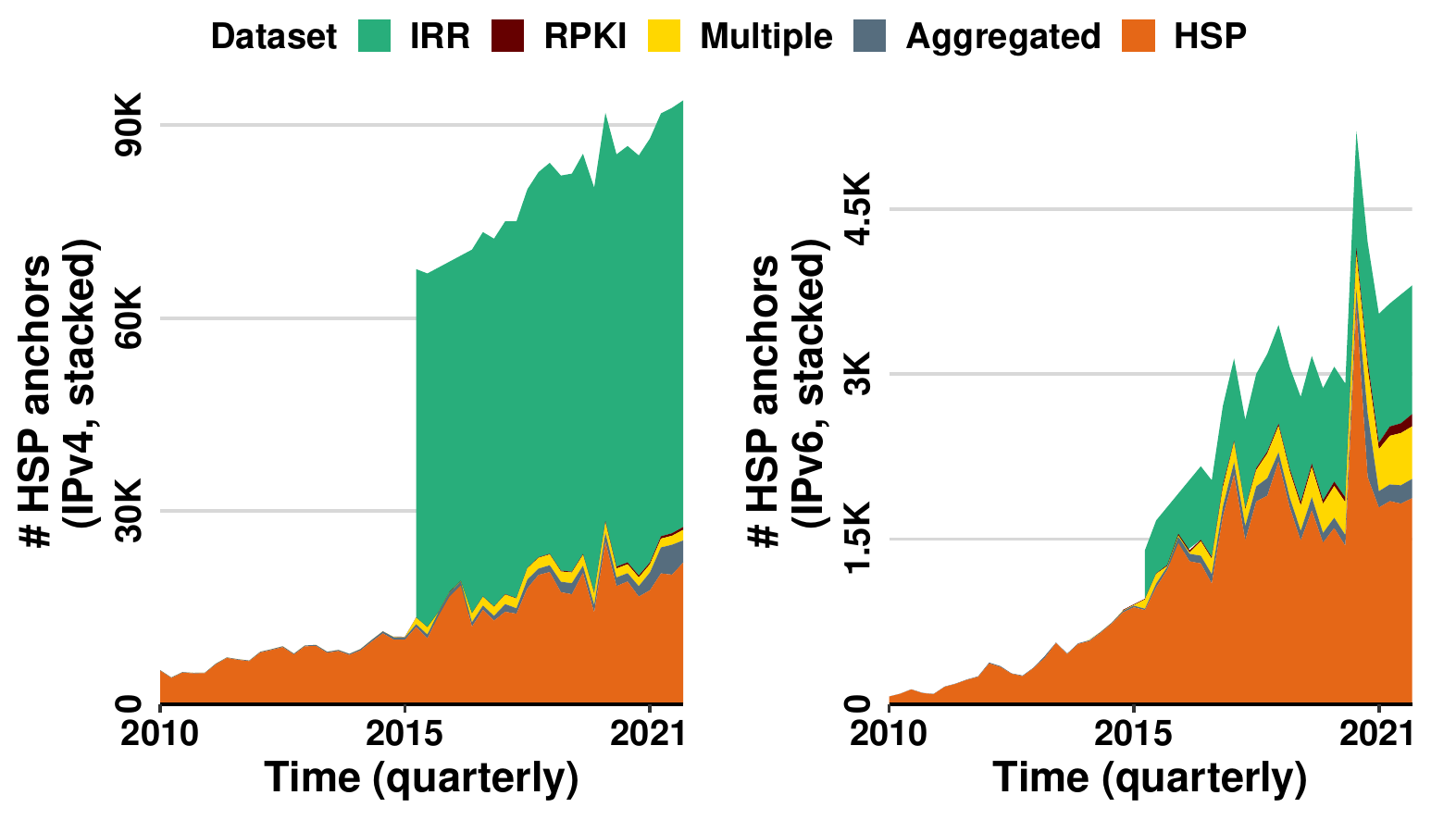}
 \caption{HSP anchors across data sets
 }\label{fig:rca:msp-across-datasets}
\endminipage
\end{figure*}

\textbf{Where does HSP aggregation happen?} Now that we have a set of anchor-prefixes that are likely the result of prefix-based aggregations, we can analyze how close to the origin HSPs are aggregated. We compare the AS number in the \texttt{AGGREGATOR} field with the \texttt{AS\_PATH} and differentiate between the following cases: \one \textbf{Origin}---the origin itself performed the aggregation, \two \textbf{On-path}---an AS within the AS path that is not the origin performed the aggregation, and \three \textbf{Off-path}---some AS that does not occur in the AS path performed the aggregation\footnote{This class also includes reserved AS numbers.}. Figure \ref{fig:rca:hsp-aggr-position} shows the number of anchor prefixes in each class over time. Notably, the figure also contains the class \textbf{Multiple} that contains anchor prefixes for which there are multiple paths with inconsistent classes. We observe that the the vast majority of anchors are actually aggregated at the origin with only few hundreds of anchors being aggregated on-path. Origin and off-path (especially \texttt{AGGREGATOR} fields with private ASNs) aggregation often occurs due to the use of BGP confederations~\cite{Juniper2021Confeder,Cisco2021Confeder} where the AS is internally split into multiple private sub-ASes. Depending on how an AS border router handles the aggregation of internal confederation routes, it might either correctly set the external AS number or leak the internal confederation AS Number in the \texttt{AS\_PATH} or \texttt{AGGREGATOR} attribute. Notably, those HSP routes are likely not available to other ASes (including neighbors of the origin).

\textbf{Projected actual usage.} While our IRR snapshots produced actual HSPs, our final prefix-aggregation and ROAs only produced a list of anchor-prefixes that is likely to contain HSPs. Therefore, we decided to analyze the potential extent of HSP usage on the basis of anchor-prefixes. Figure \ref{fig:rca:msp-across-datasets} shows the number of IPv4 (left) and IPv6 (right) anchor-prefixes per data set (stacked) over time. Notably, the aggregated class only contains on-path aggregated anchor prefixes and the RPKI class only contains anchor prefixes for explicit HSP ROAs. The ``multiple`` class covers those entries that are visible via multiple data sources. We observe that the current route collector infrastructure misses roughly one-third of the of the anchor prefixes that potentially contain HSPs. We further observe a less noisy, linear increase in the number of anchor prefix for which HSPs are visible compared to the raw count of visible HSPs. Notably, some part of this increase can potentially be accounted to the increasing numbers of route collectors and route collector peers over time.

\textbf{Who uses hyper-specific prefixes?}
We leverage the ``AS Classification Inferences'' dataset described in ASDB~\cite{ziv2021asdb} to classify ASes as Content, Education, Hypergiant, ISP (Stub), ISP (Transit), Tier 1, and Others.
\Cref{fig:rca:hsp_origins_class} compares the classes of all BGP-visible ASes (left) to \hspS origin ASes (right) over time.
We find that in contrast to all origin ASes, \hspS origins are more likely to be ISP (Transit) ASes.
Interestingly, the majority of Tier~1 ASes is also originating \hspsS.
During the period of January 2019 until October 2021, we identify between 12 and 15 of the total 19 Tier~1's as \hspS origins.
In contrast to the high share of Tier~1 \hspS origins, we find that most hypergiants do not originate HSPs.

\section{Real-World Experiment}
\label{appendix:testbed}
\textbf{Does BGP reflect control plane reachability?} Finally, we want to understand how much the lack of additional BGP vantage points impacts our observations on reachability. Hence, we configure a real-world experiment using the PEERING testbed~\cite{schlinker19peering} in which we announce an anchor prefix as well as multiple hyper-specific prefixes. Once those prefixes have converged, we run traceroutes from RIPE Atlas~\cite{ripeatlas} probes and compare their resulting paths to those visible at route collectors.
\begin{figure*}[!htb]
	\minipage[t]{0.66\textwidth}
	    \includegraphics[width=\linewidth]{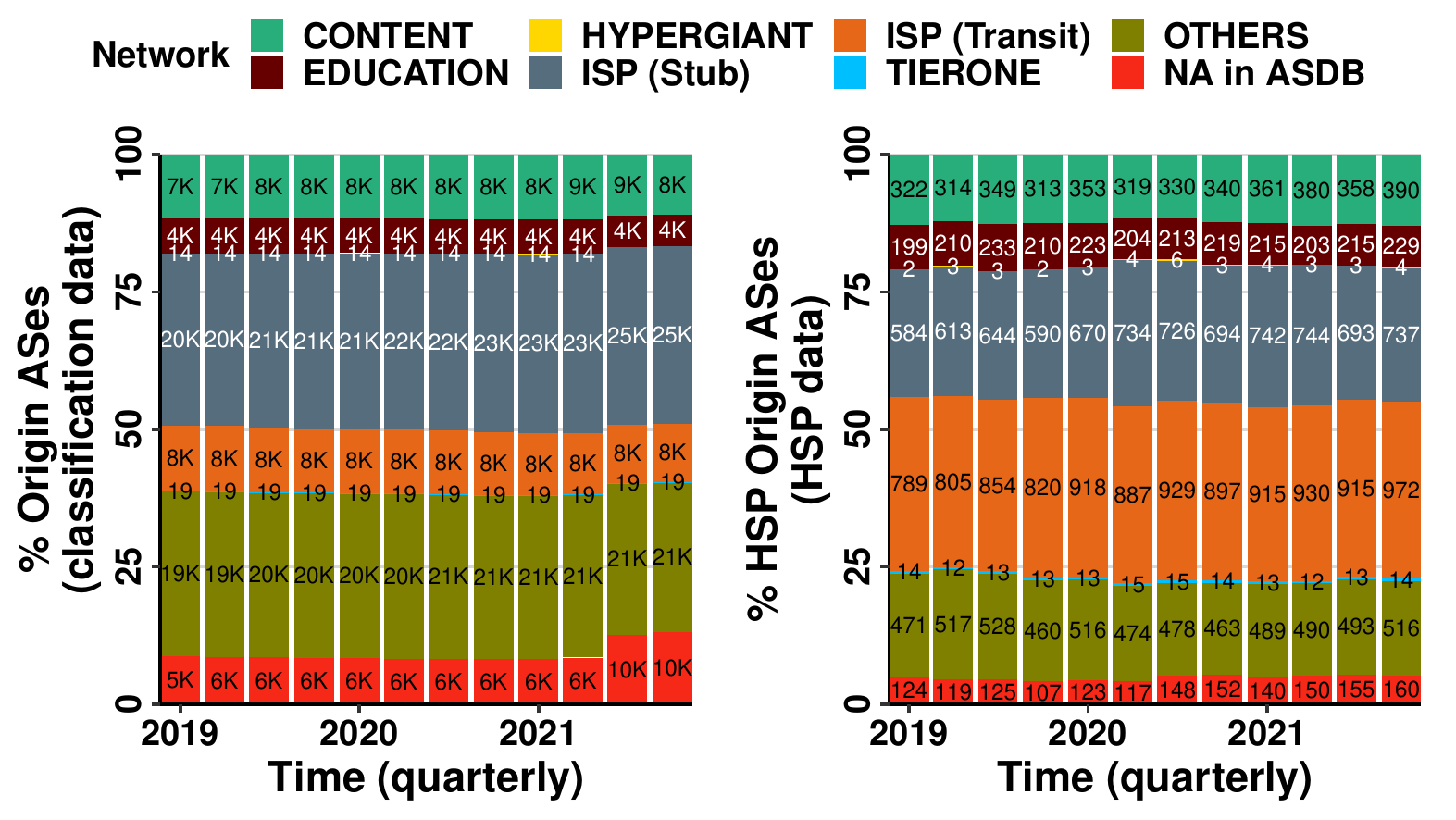}
	    \caption{HSP origin AS classification over time.
	    }\label{fig:rca:hsp_origins_class}
	\endminipage
	\hfill
	\minipage[t]{0.32\textwidth}
    \includegraphics[width=\linewidth]{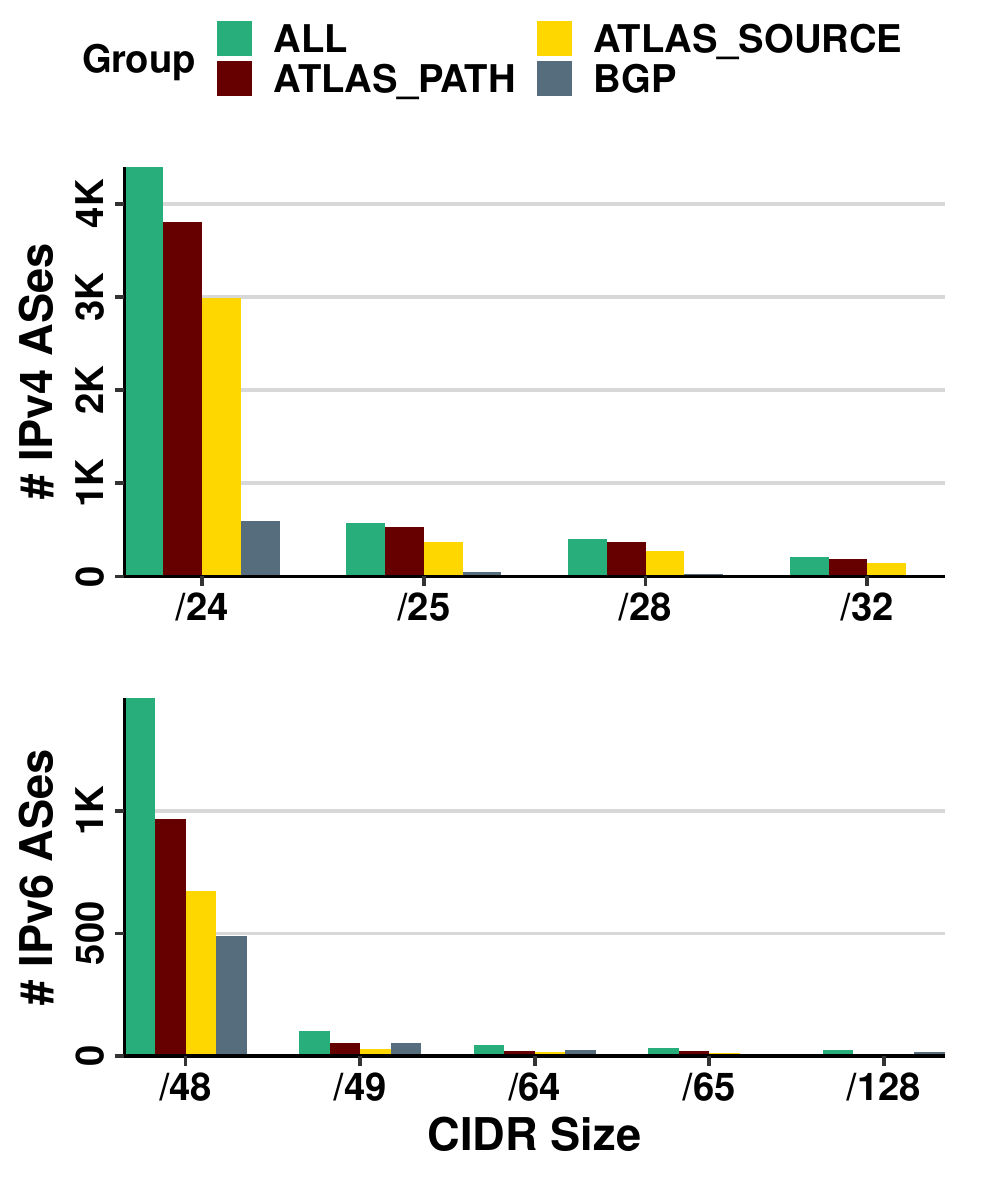}
    \caption{PEERING testbed propagation results
    }\label{fig:peering}
  \endminipage

\end{figure*}

\textit{Vantage points \& resources.} The PEERING testbed allocates Internet resources (specifically, IPv4/IPv6 address space and AS numbers) to its users based on approved experiment proposals. Once allocated, users can announce those resources via the testbed's infrastructure. Given that the PEERING testbed strongly relies on third party resources (e.g., for hosting infrastructure), announcements must be designed carefully to not cause trouble or irritation for other network operators. For our experiment we use the address ranges 184.164.240.0/23 and 2804:269c:4::/46. More specifically, we utilize 184.164.240.0/24 and 2804:269c:4::/48 as anchor prefixes (i.e., they represent our control group) and announce HSPs only from the remaining address space\footnote{In particular, we announce 184.164.241.0/25, 184.164.241.128/28, 184.164.241.255/32, 2804:269c:5::/49, 2804:269c:6::/64, 2804:269c:6:8000::/65, and 2804:269c:7::/128.}.

RIPE Atlas~\cite{ripeatlas} is a measurement platform with probing devices (henceforth called probes) all over the world. To maximize probing coverage and minimize probing load, we choose at most one probe per AS. To reduce the likelihood of probe outage, we select only probes that are not tagged with system-problematic tags\footnote{tags: \texttt{system-flash-drive-filesystem-corrupted}, \texttt{system-v1}, \texttt{system-no-flash-drive}, \texttt{system-flash-drive-bad-or-too-small}, \texttt{system-firewall-problem-suspected}, \texttt{system-trying-to-connect} \texttt{system-readonly-flash-drive}, \texttt{system-no-controller-connection}, \texttt{system-bad-firmware-signature}, \texttt{system-flakey-connection}, \texttt{system-flakey-power}, \texttt{system-flash-drive-problem-detected}, and \texttt{system-v2}}. We further validate that selected IPv4 and IPv6 probes are tagged with \texttt{system-ipv4-works}, and \texttt{system-ipv6-works}, respectively. If an AS hosts multiple probes, we prefer dual-stack probes (such that we can use a consistent probe for our IPv4 and IPv6 measurements) over anchor probes (i.e., better equipped probes) over any other probes. If we still have multiple choices, we pick the probe that is tagged with the highest stability tag (e.g., \\\texttt{system-ipv4-stable-90d}). %
Our final probe set consists of 3097 probes distributed across 2990 IPv4 and 1239 IPv6 ASes.

\textit{Experimentation environment.} The PEERING testbed currently has a total of 180 IPv4 and 152 IPv6 neighboring ASes. Yet, most neighbors do not support/redistribute HSPs. We identify supportive neighbors by iteratively announcing a /25 or /49 prefix from our allocated address space through each neighbor and analyzing the resulting update stream from RIPE RIS and Routeviews. Since, at this point, we only care about a ``life sign`` (i.e., whether or not \textit{any} update was received) rather than full convergence, we adopt a short announcement cycle: We announce a prefix at the beginning of every full hour and withdraw it 30 minutes later\footnote{These experiments ran between the May 1, 2021 and the May 3, 2021.}. We identify a set of 8 IPv4 and 9 IPv6 neighboring ASes that redistribute HSPs. Notably, those ASes are distributed across 4 and 3 geographically separate Points of Presence.

\textit{Technical realization.} Throughout May 21 and 22, 2021, we announce /24, /25, /28, and /32 IPv4 prefixes and /48, /49, /64, /65, /128 prefixes through a single neighbor at the beginning of every even hour. After the announcement, we wait 40 minutes to allow the prefix to converge\footnote{During previous experiments we observed that usually the 95th percentile of updates reach the collector peers already in the first 15 minutes.}. After those 40 minutes, we run active measurements for 10 minutes, and then withdraw the prefixes again. Notably, we choose 70 minutes between a withdrawal and the next announcement on purpose such that we out-wait the expiration of potential Route Flap Damping hold-down timers, which have been shown to usually expire after 60 or less minutes~\cite{gray2020bgp}. %

During our 10 minute active measurement period, we run paris-traceroutes from all probes towards either the network address or the first non-network address of all prefixes (which are configured to be pingable). To reduce the dependence of our results on the underlying protocol, we simultaneously issue ICMP, TCP, and UDP probing. To keep the induced load for the RIPE Atlas platform as well as for the peering testbed manageable, we reduce the number of probing packets used per per hop by paris-traceroute from RIPE ATLAS' default of 3 packets to one packet. Notably, as the resulting load still exceeds the default limitations (e.g., for measurement results per day) for a single RIPE Atlas account, we coordinate our probing efforts with the RIPE Atlas team who generously raised the limits for our experiments.

We map traceroutes to AS Paths using the state-of-the-art mapping tool bdrmapit~\cite{marder2018pushing}. As bdrmapit requires a large corpus of traceroutes as input to perform well, we use traceroute data from CAIDA's IPv4 Prefix-Probing data set~\cite{Caida2021Routed24}%
, CAIDA's IPv4 Routed /24 Topology Dataset~\cite{Caida2021Routed24}, CAIDA's IPv4 Routed /48 Topology Dataset~\cite{Caida2021routed48}, and RIPE's hourly archives of Atlas traceroutes~\cite{RIPE2021dumps} between May 17, 00:00 and May 24, 00:00. For all the other inputs (e.g., prefix-to-origin mappings or business relationship inferences) we use recent snapshots from the recommended data sources. Finally, we use bdrmapit's output to map our successful (i.e., only those that actually reached the respective target host) traceroutes to AS paths.

\textit{Comparison.} Figure \ref{fig:peering} compares the the number of ASes (aggregated over all iterations) that \one hosted Atlas probes that reached the target (ATLAS\_SOURCE, yellow), \two appeared along the path between ATLAS\_SOURCE ASes and the Peering Testbed (ATLAS\_PATH, dark red), \three are visible from route collector peers (BGP, gray). The most drastic observation is that hyper-specific prefixes see a very sharp drop in reachability. Even the best performing CIDR size, /25, only reached \textasciitilde15~\% of of the ASes that are reached by its respective anchor prefix. Especially for IPv6 we observe that most PEERING neighbors redistribute our prefixes (including the anchor prefix) only towards their customers, hence, some of our Atlas probes are unable to reach the peering testbed even for the anchor prefix.
We further find that the more-specific the prefix gets, the less likely it propagates. This finding is interesting as most recommended filtering guides~\cite{noction2020prefix,rfc7454,nlnog2020filtering,Gert2013transit,MANRS2021Filter} treat all hyper-specific CIDR sizes equally. Our third observation is that the reachability reflected by route collector peers substantially underestimates data plane reachability. While we are able to observe approximately a third of the total ASes for our /48 prefix via BGP, this fraction lies at around 14~\% for our /24 prefix.

\section{Filtering Pipeline}
\label{appendix:filtering}
When an AS peers with a Route Collector, the router that feeds the collector may provide all routes that are not removed during (or before) egress filtering. Hence, misconfigured egress filters can lead to misinterpretations. For our analysis, we filter out HSPs which are originated by feeder AS directly connected to a route collector. However, we use the HSP if it has been propagated to at least 2 AS hops, including feeder AS.  In addition, we filter all private, reserved, multicast, and experimental IP prefixes. Furthermore, we also filter prefixes originated by a private AS. Finally, we remove the HSPs we identify as outliers during the data cleaning process. \Cref{app:list_filtered_hsps} provides detail information on HSPs we have filtered out.

\begin{table*}[t!]
	\section{Applied Data Isolation Rules}\label{app:list_filtered_hsps}
		\begin{tabular}{llll}
            \toprule
		 	Timeframe		& Filter name			& Filter Details	& Reason\\
            \midrule
		 entire period 	&  Private Origin ASes	& Origin AS number from	&  private IPv4 ranges.\\
		  	& 	2 Bytes			& 	64512 to 65534		& \\
		 entire period 	&  Private Origin ASes	& Origin AS number from 	&  private IPv4 ranges.\\
		 	& 		4 Bytes					& 4200000000 to 4294967294	& \\
		 entire period 	&  Private IPs	& IPv4 Private IP ranges	&  private IPv4 ranges.\\
		  	& 							& 										  	& \\

		 entire period 	&  Class D and E	& IPv4 Prefixes > 223.x.x.x	& IPv4 multicast  and class E IP ranges.\\
		  	& 							& 										  	& \\

		 entire period 	&  Abnormal Prefixes	& for IPv4 prefix > /32	&  abnormal IPv4 prefixes.\\
		 	& 							& 			for IPv6 prefix > /128 & abnormal IPv6 prefixes\\

	 entire period 	&  No Origin	& Routes having no origin AS 	&   AS-internal routes.\\
	  	& Internal	&  Feeder AS is the Origin AS	 & \\

	2015/10/01-07 	&  IPv4 Noisy Origins &  Origin AS == 9498 	&  routes from particular origin AS.$\textcolor{red}{\star}$\\

 2016/10/01-07 	&  IPv4  Noisy Origins &  Origin AS == 36937 	&  routes from particular origin AS.$\textcolor{red}{\star}$\\

 2017/04/01-07 	&  IPv4 Noisy Origins &  Origin AS == 9498 	&  routes from particular origin AS.$\textcolor{red}{\star}$\\
2019/07/01-07 	&  IPv4  Noisy Origins &  Origin AS 7122 	&  routes from particular origin AS.$\textcolor{red}{\star}$\\

 entire period 	&  IPv4 Noisy Origins &  Origin AS 12400 	&  routes from particular origin AS.$\textcolor{red}{\star}$\\

 2016/07/01-07 	&  IPv4 Noisy Peer AS &  Peer AS 35908 	&  routes from particular peer AS.$\textcolor{red}{\star}$\\

 2017/01/01-07 	&  IPv4 Noisy Peer AS &  Peer AS 60924 and 27630 	&  routes from particular peer AS.$\textcolor{red}{\star}$\\

 2017/10/01-07 	&  IPv4 Noisy Peer AS &  Peer AS 37497 	&  routes from particular peer AS.$\textcolor{red}{\star}$\\

 2018/10/01-07 	&  IPv4 Noisy Peer AS &  Peer AS 14361 	&  routes from particular peer AS$\textcolor{red}{\star}$\\

 2019/01/01-07 	&  IPv4 Noisy Peer AS &  Peer AS 262757 	&  routes from particular peer AS.$\textcolor{red}{\star}$\\

 2020/04/01-07 	&  IPv4 Noisy Peer AS &  Peer AS 268430 	&  routes from particular peer AS.$\textcolor{red}{\star}$\\

 2021/04-07/01-07 	&  IPv4 Noisy Peer AS &  Peer AS 398465 	&  routes from particular peer AS.$\textcolor{red}{\star}$\\
 2021/01-10/01-07 	&  IPv4 Noisy Peer AS &  Peer AS 203125 	&  routes from particular peer AS.$\textcolor{red}{\star}$\\

 2020/04-07/01-07 	&  IPv4 Noisy Peer AS &  Peer AS 268430 	&  routes from particular peer AS.$\textcolor{red}{\star}$\\

 entire period 	&  IPv6 Noisy Origins &  Origin AS 4761 	&  routes from particular origin AS.$\textcolor{red}{\star}$\\

 2017/07/01-07  	&  IPv6 Noisy Origins &  Origin AS 17451 and 45899	&  routes from particular origin AS.$\textcolor{red}{\star}$\\

 2019/04/01-07  	&  IPv6 Noisy Origins &  Origin AS 7713	&  routes from particular origin AS.$\textcolor{red}{\star}$\\

 2021/07/01-07  	&  IPv6 Noisy Origins &  Origin AS 8100	&  routes from particular origin AS.$\textcolor{red}{\star}$\\

 2018/07/01-07 	&  IPv6 Noisy Peer AS &  Peer AS 199036 	&  routes from particular peer AS.$\textcolor{red}{\star}$\\
 \bottomrule
		\end{tabular}
		\caption{\label{tab:filtering} Applied filtering and isolation rules. $\textcolor{red}{\star}$: these ASes contributed/announced either \one an extraordinary high number of HSPs (i.e., 100 or more times higher than in other snapshots) or \two HSPs in an extraordinary high number of anchor prefixes for a limited time.}
\end{table*}
 
\end{document}